\documentclass{PoS}
\pdfoutput=1

\usepackage{amsmath,amssymb,amsthm,amsfonts}
\usepackage{graphicx,rotate}
\usepackage{subfigure}

\title{Yang-Mills Theory at Non-Vanishing Temperature}

\ShortTitle{Yang-Mills Theory at Non-Vanishing Temperature}

\author{\speaker{Leonard Fister}\\
        Institut f\"ur Theoretische Physik, 
Universit\"at Heidelberg, 
Philosophenweg 16, 69120 Heidelberg, Germany\\
        E-mail: \email{l.fister@thphys.uni-heidelberg.de}}

\author{Jan Martin Pawlowski\\
        Institut f\"ur Theoretische Physik, 
Universit\"at Heidelberg, 
Philosophenweg 16, 69120 Heidelberg, Germany\\
        E-mail: \email{j.pawlowski@thphys.uni-heidelberg.de}}

      \abstract{We compute ghost and gluon propagators of Yang-Mills
        theory in the Landau gauge at non-vanishing temperature within
        a functional renormalisation group setting. We construct
        purely thermal flows, that project onto thermal fluctuations
        only. For temperatures and momenta above the
        confinement-deconfinement temperature $T_c$ the electric
        propagator shows a thermal suppression due to Debye
        screening. The magnetic gluon propagator shows a thermal
        scaling and tends towards the three-dimensional one. In this
        region both propagators match the lattice propagators. The
        thermal scaling is also reflected in the infrared suppression
        of the ghost-gluon vertex. For temperatures below $T_c$ the
        electric propagator shows an enhancement which is in
        qualitative agreement with the lattice behaviour.}

      \FullConference{International Workshop on QCD Green's Functions, Confinement and Phenomenology,\\
        September 05-09, 2011\\
        Trento Italy}

\begin{document}


\section{Introduction}
In the last decades the phase diagram of quantum chromodynamics (QCD)
has been a very active field of research. For small quark chemical
potential, but non-vanishing temperature enormous progress has been
achieved, which is mainly due to the fact that both, first principle continuum
methods, see e.g.
\cite{Pawlowski:2010ht,Binosi:2009qm,Fischer:2006ub,Alkofer:2000wg,Roberts:2000aa},
as well as lattice QCD, see e.g.\
\cite{Philipsen:2011zx,Borsanyi:2010cj,Bazavov:2010sb}, have been
applied very successfully. Nonetheless, even at vanishing density,
some questions still remain open, in particular in the vicinity of the
chiral and confinement-deconfinement phase transitions. This concerns
for example the trace anomaly in this region. Indeed, for temperatures
around the deconfinement-confinement phase transition $T\lesssim 3T_c$
continuum methods so far lack quantitative precision. This temperature
regime is governed by fully non-perturbative physics, and requires the
application of non-perturbative techniques.

In continuum approaches to QCD, it is mainly the glue sector which
requires the most effort, in particular conceptually but also
technically. In the past two decades much progress has been made in
order to understand Yang-Mills theory in the Landau gauge at
vanishing temperature, leading to quantitative precision for
correlation functions, see \cite{Fischer:2008uz}. At finite
temperature, however, apart from conceptual issues, the same
quantitative precision is lacking. 

In the present contribution we report on work published in \cite{finiteTYM}. We aim at a quantitative computation of
the Yang-Mills thermal propagators, for a review on thermal gluons see
\cite{Maas:2011se}. Here we utilise the functional renormalisation
group (FRG), for QCD-related reviews see
\cite{Pawlowski:2010ht,Litim:1998yn,Litim:1998nf,Berges:2000ew,Pawlowski:2005xe,Gies:2006wv,Schaefer:2006sr,Braun:2011pp}.
The FRG incorporates non-perturbative effects by a successive
integration of fluctuations related to a given momentum scale, and
hence is applicable to all temperatures. In the presented work we will
focus on the pure Yang-Mills part of QCD, which has so far been
studied in the framework of functional methods by the help of
Dyson-Schwinger equations \cite{Maas:2011se, Gruter:2004bb, Cucchieri:2007ta}. We
compute fully dressed, non-perturbative ghost and gluon propagators,
as well as the ghost-gluon vertex at finite temperature. The electric
propagator shows the typical thermal screening above the
confinement-deconfinement temperature $T_c$ in quantitative agreement
with the lattice results, see
\cite{Maas:2011se,Cucchieri:2007ta,Fischer:2010fx,Bornyakov:2011jm,Aouane:2011fv,Maas:2011ez,Cucchieri:2011di}. It
also shows an enhancement below $T_c$ which is qualitatively similar
to the lattice findings. The magnetic propagator shows thermal scaling
and tends towards the three-dimensional one for large temperatures. As for the electric propagator, it agrees well with the lattice.

\section{Yang-Mills Green Functions with the Functional Renormalisation Group}
The FRG is an exact method that is derived directly from the
functional integral. It constitutes a renormalisation group equation
\cite{Wetterich:1992yh} for the scale-dependent effective action
$\Gamma_k$, which interpolates between the classical action in the
ultraviolet and the full effective action in the infrared by variation
of an infrared cutoff, which suppresses fluctuations with momenta
smaller than the infrared scale $k$. In the limit of $k\rightarrow0$
one is left with the full quantum effective action.

\begin{figure}[t]
\begin{center}
\includegraphics[width=.4\columnwidth]{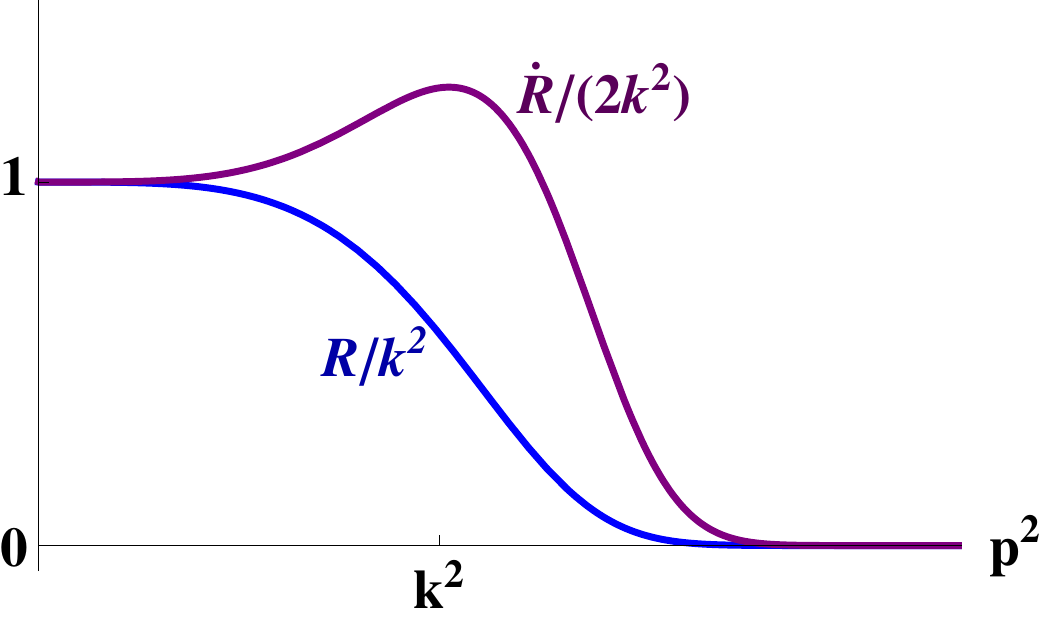}
\caption{Regulator $R(p^2)$ and its (logarithmic) cut-off scale derivative $\partial_t
  R(p^2)=\dot R(p^2)$.}
\label{fig:R}
\end{center}
\end{figure}

\begin{figure}[b]
\begin{center}
\includegraphics[width=.42\columnwidth]{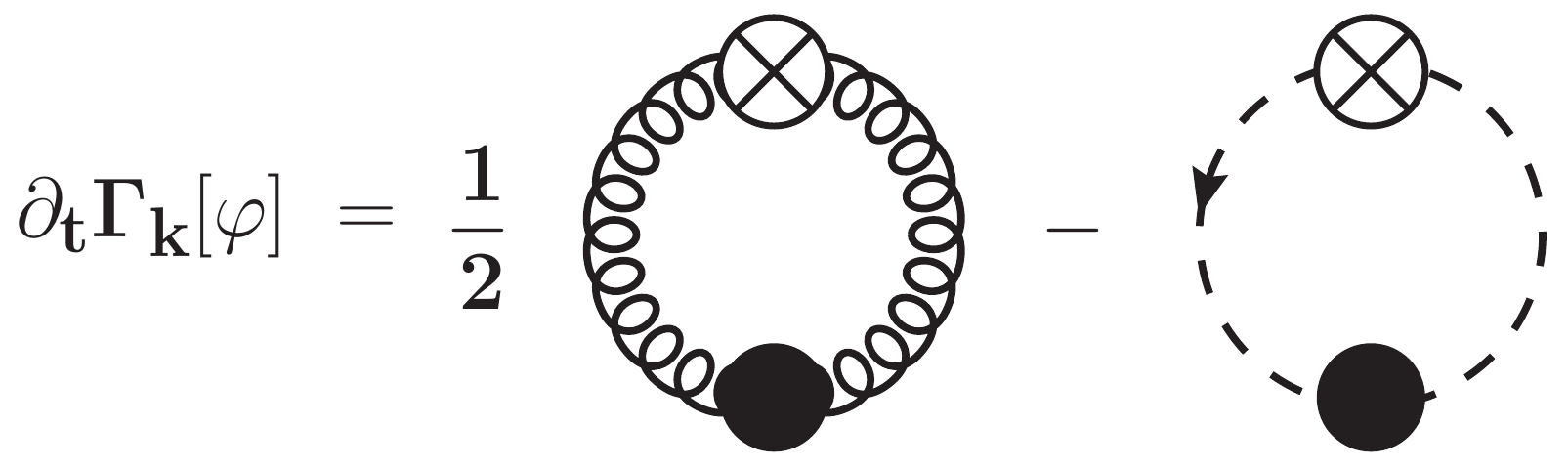}
\caption{Functional flow for the effective action. Lines with filled
  circles denote fully dressed field dependent propagators
  $G\!\left[\!\varphi\!\right] $. Crossed circles denote the regulator
  insertion $\partial_t R_k$.}
\end{center}
\label{fig:funflow}
\end{figure}

\begin{figure}[h]
\begin{center}
\includegraphics[width=.85\columnwidth]{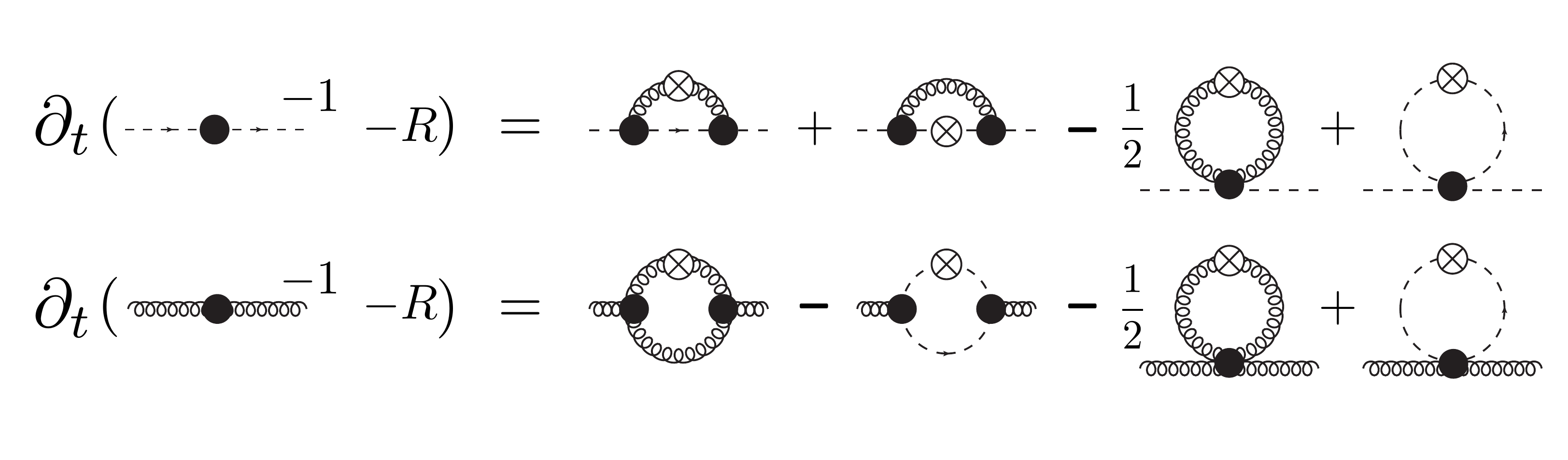}
\caption{Flow equations for the ghost and gluon propagators.
  Vertices with filled circles denote fully dressed vertices. All
  propagators are fully dressed, the filled circles for the internal
  ones have been dropped for the sake of clarity.  Crossed circles
  denote the regulator insertion $\partial_t R_k$.}
\label{fig:YM_props}
\end{center}
\end{figure}

The classical gauge fixed $SU(N_c)$ Yang-Mills action in
Euclidean space in the Landau gauge is given by
\begin{equation}
  S=\int d^4x\left( \frac{1}{4} F_{\mu\nu}^a F_{\mu\nu}^a+
    \overline c^a\partial_{\mu} D_{\mu}^{ab} c^b \right)\, ,
\end{equation}
with the ghost fields $\overline{c}^a, c^a$. The field strength tensor
$F_{\mu\nu}^a$ and the covariant derivative $D_{\mu}^{a b}$ are defined in terms of
the gluons $A_{\mu}^a$, the coupling $g$ and the structure constants
$f^{abc}$ of the gauge group, 
\begin{eqnarray}
  F^a_{\mu\nu}=\partial_{\mu} A^a_{\nu}-\partial_{\nu} A_{\mu}^a+gf^{abc} A_{\mu}^b 
  A_{\nu}^c, 
  \qquad  D_{\mu}^{ab}=\delta^{ab}\partial_{\mu}+g f^{acb} A_{\mu}^c\,.
\end{eqnarray} 
In the FRG-framework, infrared fluctuations below the scale $k$ are
suppressed by a modification of the propagator in the classical action
of the form $S \to S+\Delta S_k$,
with
\begin{equation}
\Delta S_k=\int \frac{d^4p}{\left(  2 \pi \right)^4} 
 \frac{1}{2}  A^{a}_{\mu}\, R_{\mu\nu}^{ab}(p^2) \, A^{b}_{\nu}  
+\int \frac{d^4p}{\left(  2 \pi \right)^4}  \bar c^{a}\, R^{ab}(p^2)\, c^{b}\, .
\end{equation}
The regulator functions $R_{\mu\nu}^{ab}$ and $R^{ab}$ carry the same
Lorentz and colour structure as the corresponding field propagators
and are proportional to the dimensionless shape function $r$ such that
schematically $R(p^2,k^2)\sim p^2 r(p^2/k^2)$. Note that the ghost
regulator is negative related to the unphysical dispersion of the
ghost. In the following we use regulators of the form given in
Fig.~\ref{fig:R}.
The flow equation for the scale-dependent Yang-Mills effective action
$\Gamma_k$ at finite temperature is given by
\begin{equation}  
\partial_t \Gamma_{k}[\varphi] =  
  \frac{1}{2} \textnormal{Tr}\, G[\varphi](p,p)
  \cdot {\partial_t} R_k(p)\,. 
\label{eq:funflow}\end{equation}
The trace stands for summation over species of fields with a relative
minus sign for fermions, a summation over Lorentz and colour indices,
and the integration/summation of the loop momenta. We also introduced
the compact notation $\varphi = (A,c,\overline{c})$. In
eq.~\!(\ref{eq:funflow}) $G\!\left[\varphi  \right]$ is the short
notation for the fully dressed field dependent propagator. At
imaginary time in the Matsubara formalism the integration measure at
finite temperature turns the integration over the temporal momentum
component into a sum over Matsubara frequencies,
\begin{equation}\label{eq:matsubaras}
  \int\!\! \frac{d^4p}{(2\pi)^4}\ \rightarrow \ T 
  \sum_{n\in \mathbb{Z}}\int\!\! \frac{d^3 p}{(2 \pi)^3}\,,
  \quad {\rm with}\quad  p_0=2 \pi T n\,.
\end{equation} 
The functional flow equation for the effective action
(\ref{eq:funflow}) can be illustrated in a diagrammatic form, which is
given in Fig.~\ref{fig:funflow}. The minus sign in front of the ghost
loops originates in the fermionic nature of the ghost. 
The flow equations for the propagators can be obtained by taking
functional derivatives with respect to two gluons for the gluon
propagator, and with respect to a ghost and an antighost for the
ghost-propagator. These equations for the Yang-Mills two-point
functions are depicted in Fig.~\ref{fig:YM_props}. They only contain
one-loop terms in full propagators and vertices. We emphasise that the
flow equation depicted in Fig.~\ref{fig:YM_props} is exact, no higher
loop terms are missing. This originates in the one-loop form of
(\ref{eq:funflow}), see also Fig.~\ref{fig:funflow}.

\section{Thermal Fluctuations}
\begin{figure}[t]
\begin{center}
\includegraphics[width=.55\columnwidth]{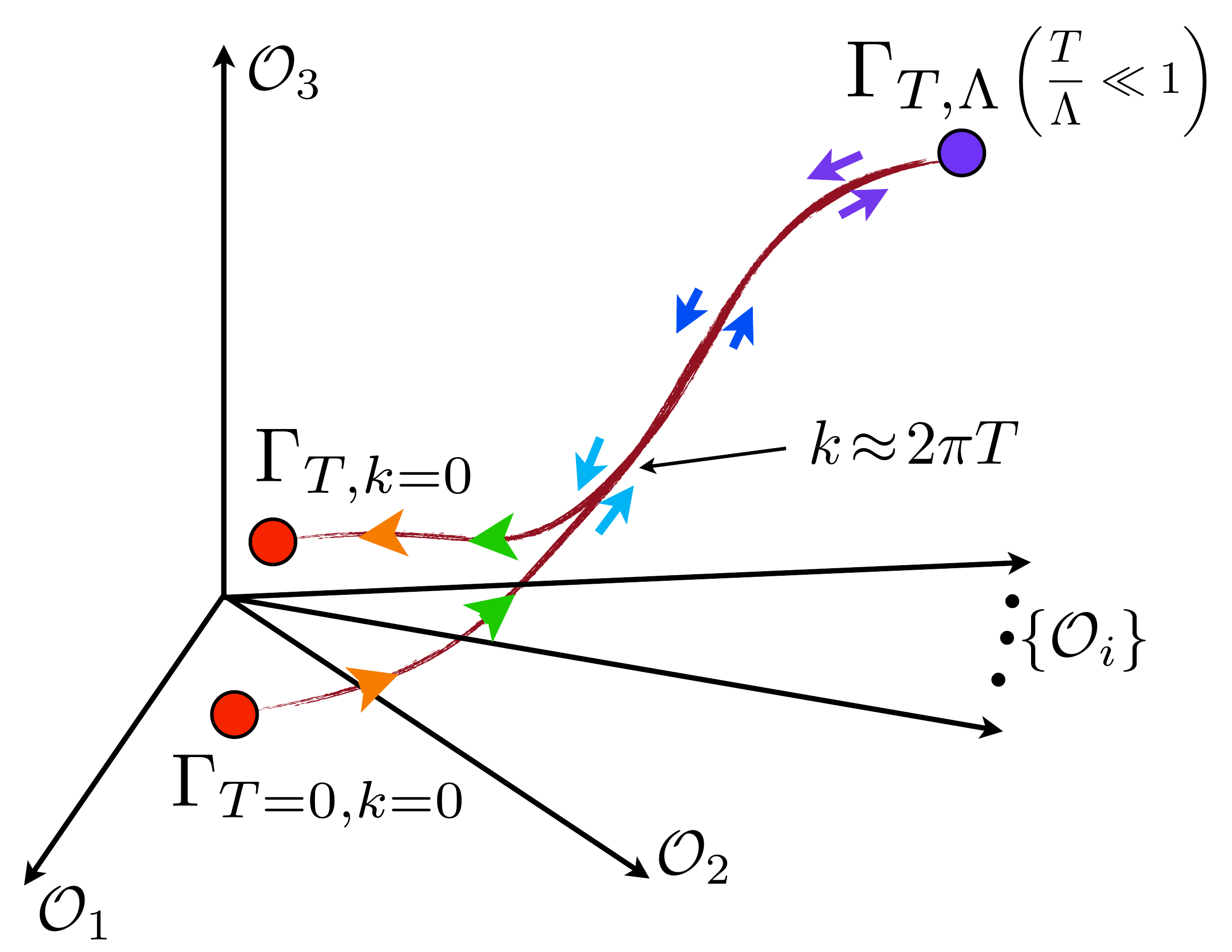}
\caption{Flow at vanishing temperature from $k=0$ to $k=\Lambda$, and
  flow at $T\neq 0$ from $k=\Lambda$ to $k=0$ with $T/\Lambda\ll
  1$. The flow is described in theory space, and the axes label
  (orthogonal) couplings/observables $\mathcal{O}_i$ which serve as
  expansion coefficients of the effective action,
  e.g. $\mathcal{O}_1=\Gamma^{(2)}(p=0)$. The flows start to deviate
  at $k\approx 2 \pi T$.}
 \label{fig:thermalflow}
\end{center}
\end{figure}

The integration of the flow equation (\ref{eq:funflow}) involves both,
quantum as well as thermal fluctuations. However, the treatment
of the thermal fluctuations can be done separately via the help of
purely thermal flows, which are constructed as the difference of the
flows at finite and zero temperature, 
\cite{Litim:1998nf,Litim:2006ag},
\begin{eqnarray}\label{eq:thermalflucs} 
&  \partial_t \Delta\Gamma_{T,k}[\phi]=\left.\frac{1}{2} \textnormal{Tr}\, G[\phi]
    \cdot \partial_t R\right|_{T}- 
  \left.\frac{1}{2} \textnormal{Tr}\, G[\phi]\cdot \partial_t R\right|_{T=0}\,.
\end{eqnarray}
The quantity
$\Delta\Gamma_{T,k}[\phi]=\left.\Gamma_{k}[\phi]\right|_{T}-
\left.\Gamma_{k}[\phi]\right|_{T=0}$
accounts for the difference between the effective action at finite and
zero temperature. For the implementation of this procedure the vacuum
physics is used as the starting point. A given set of correlation
functions at vanishing temperature and scale $k=0$ is integrated by
the help of the flow (\ref{eq:funflow}) up to a scale $\Lambda$, which
is chosen such, that all thermal fluctuations are suppressed for the
considered temperature $T$. Thus at this scale the correlation
functions are not altered at leading order by switching on the
temperature. With the flow at finite temperature we can flow down to
$k=0$ again, this time integrating thermal fluctuations as well. Note
that within this procedure only the difference $\partial_t \Delta
\Gamma_{T,k}$, i.e. only thermal effects, are sensitive to the chosen
approximation.

\begin{figure}[t]
\begin{center}
\includegraphics[width=.75\columnwidth]{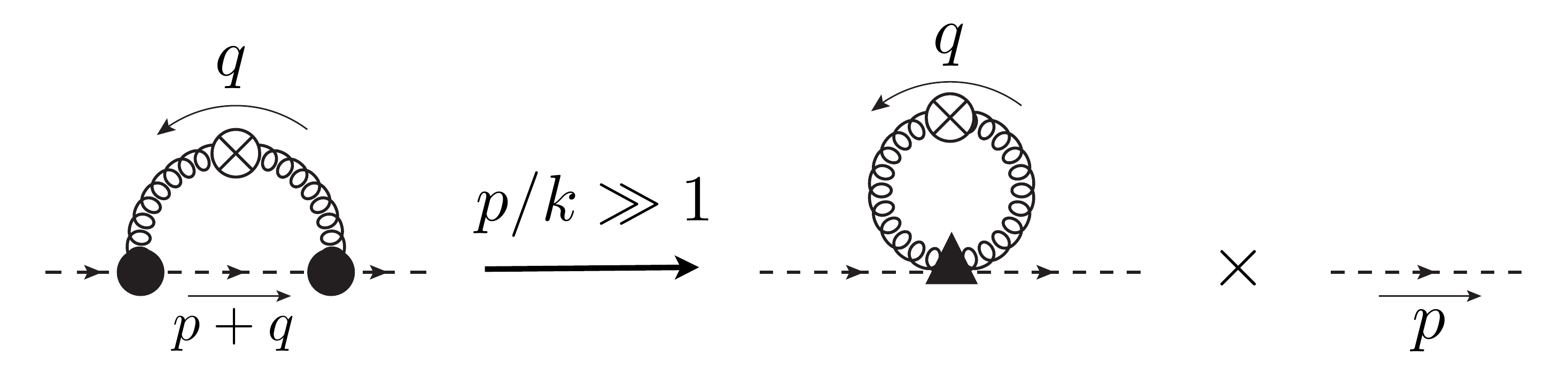}
\caption[]{Factorisation in leading order for large momenta for the
  first diagram (cutted gluon line) in the flow of the (inverse) ghost
  propagator given in Fig.~\ref{fig:YM_props}. The triangle stands for the
  product of the two vertices at $q=0$ and reads $-p_\mu p_{\mu'} f^{acd} f^{b c' d'}$. 
  }
 \label{fig:factorisation}
\end{center}
\end{figure}

The idea of the thermal flow is illustrated in a heuristic plot in
Fig.~\ref{fig:thermalflow}. However, note that for increasing
renormalisation group scale $k$ the decay
$\Delta\Gamma_{T,k}^{(n)}(p_1,...,p_n)\rightarrow0$ is exponential 
with exponent $k/T$ but only polynomially in $p_n^2/k^2$, which is an
important issue for numerical accuracy of thermodynamic
quantities. This polynomial suppression is due to the fact, that only
the loop-momenta are constrained by the regulator $q\lesssim k$, but
the external momenta $p$ are not limited. For large momenta $p \gg k$
the flow factorises at leading order, see
Fig.~\ref{fig:factorisation}, which weakens the locality of the flow.
However, the locality of a flow can be restored by a $k$-dependent
reparametrisation of the fields, which can be done in such a way, that
the flow of the two-point functions is strictly zero for external
momenta larger than a certain scale $p>\lambda k$,
\cite{Pawlowski:2005xe,finiteTYM}. Note that the contributions stemming
from this region are not dropped, but rather stored in a
reparametrisation function of the fields. This reparametrisation must
be undone at the end of the computation. In the following we will drop
any reference to the reparametrisation of fields and refer the
interested reader to \cite{finiteTYM}.

\section{Approximation}
The flow of an arbitrary $n$-point function depends itself on
$m$-point functions with $m\leq n+2$, see e.g.\ that of the
propagators depicted in Fig.~\ref{fig:YM_props}. As a consequence, the
flow in eq.~(\ref{eq:funflow}) generates an infinite hierarchy of
coupled integro-differential equations. Within numerical applications
this system must be closed by projecting it on a finite set of flow
equations while keeping the relevant physics. In other words, the
contributing correlation functions must already comprise the relevant
physics. For the propagators we keep the full propagators and work
with self-consistent approximations to the vertices which respect the
renormalisation group properties of the vertices. Furthermore, we use
the flow equation for the ghost-gluon vertex at the symmetric point with 
$p^2=k^2$. 

In the Landau gauge the gluon is a purely transversal vector
particle. At finite temperature the heat bath singles out a preferred
rest frame defined by a time-like vector $n_\mu$. This leaves us with
one ($3d$) longitudinal and one transversal tensor structure, both 
being transverse in $4d$.  The corresponding projection operators
are given by
\begin{eqnarray}
  \Pi^T_{\mu \nu}(p)&=&\delta_{\mu \nu}-p_{\mu}p_{\nu}/p^2\,,\nonumber\\
  P^{T}_{\mu \nu}(p_0, \vec{p}) &=& \left(1-\delta_{\mu 0} \right)
\left(1-\delta_{\nu 0} \right) \left( \delta_{\mu \nu}- p_{\mu}p_{\nu}/\vec{p}^{\ 2} \right)\,,
  \nonumber\\
  P^{L}_{\mu \nu}(p_0, \vec{p}) &=& \Pi^T_{\mu \nu}(p)-P^{T}_{\mu \nu}(p_0, \vec{p}) .
  \label{eq:projections} 
\end{eqnarray}
At vanishing temperature the gluon can be parametrised with one scalar
function. At non-vanishing temperature the wave-function
renormalisations for the longitudinal propagator $Z_L$ and transverse
propagator $Z_T$ differ and have to be taken into account
separately. The ghost has no Lorenz structure and can be described fully
in terms of one scalar function also at non-vanishing
temperature. Leaving out the identity in colour-space the
parametrisation is done according to
\begin{eqnarray} 
  \left(\Gamma_{A,L}^{(2)}\right)_{\mu\nu}(p_0,\vec{p}) &=&
  Z_{L}(p_0, \vec{p})\, p^2\,P^{L}_{\mu\nu}(p_0,
  \vec{p})\,, \nonumber\\
  \left(\Gamma_{A,T}^{(2)}\right)_{\mu\nu}(p_0,\vec{p}) &=& Z_{T}(p_0, \vec{p})\, p^2\,P^{T}_{\mu\nu}(p_0,
  \vec{p})\,,
  \nonumber \\
  \Gamma_{c}^{(2)}(p_0,\vec{p})&= & Z_c(p_0, \vec{p})\,p^2\,,
 \label{eq:parahatG}
\end{eqnarray}
where we emphasise that the wave-function renormalisations depend on
the temporal momentum component $p_0$ and the spatial ones $\vec{p}$
separately.

\begin{figure}[]
\begin{center}
\includegraphics[width=.2\columnwidth]{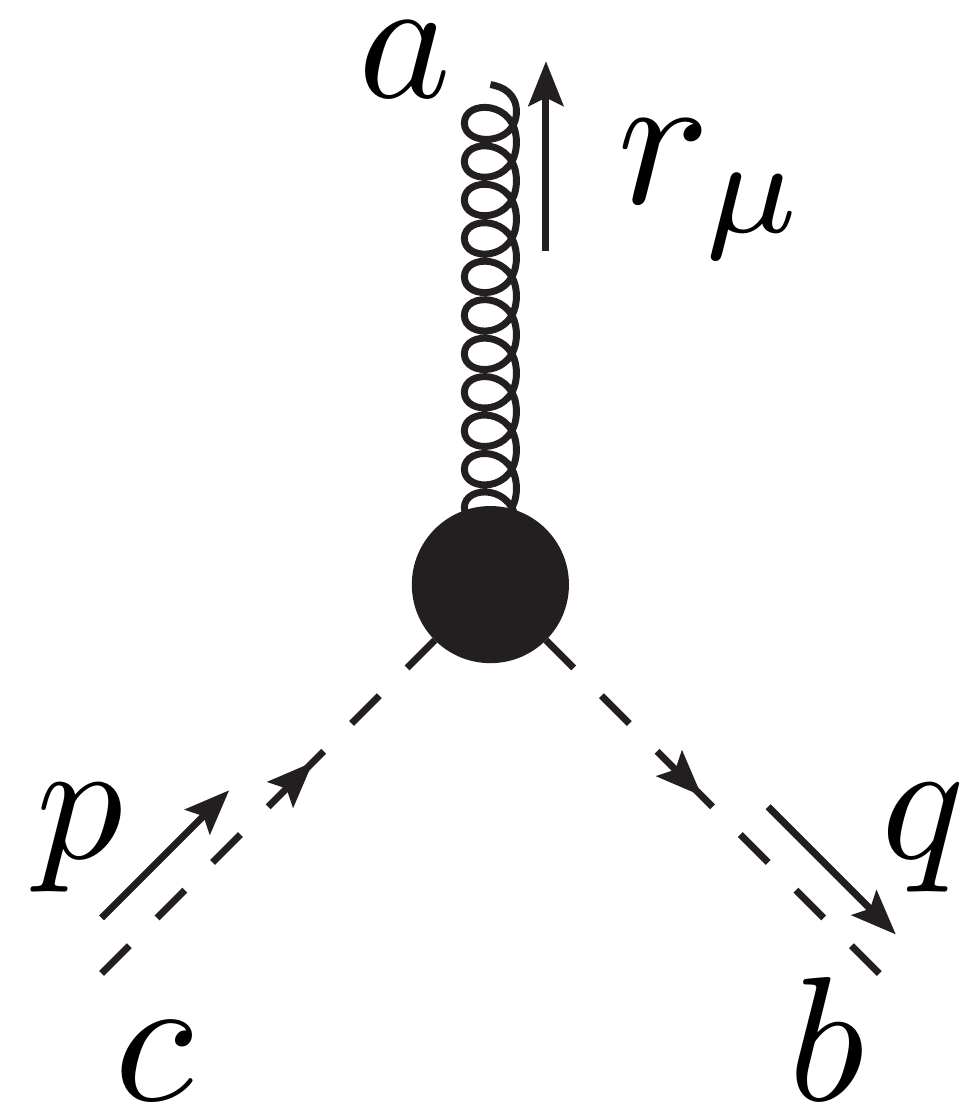}
\caption[]{Ghost-gluon vertex.}
\label{fig:cAc}
\end{center}
\end{figure}

\begin{figure}[h!]
\begin{center}
  \subfigure[Flows for the two-point functions in the given 
truncation.]{\label{fig:Trunc_props}\includegraphics[width=0.8
\textwidth]{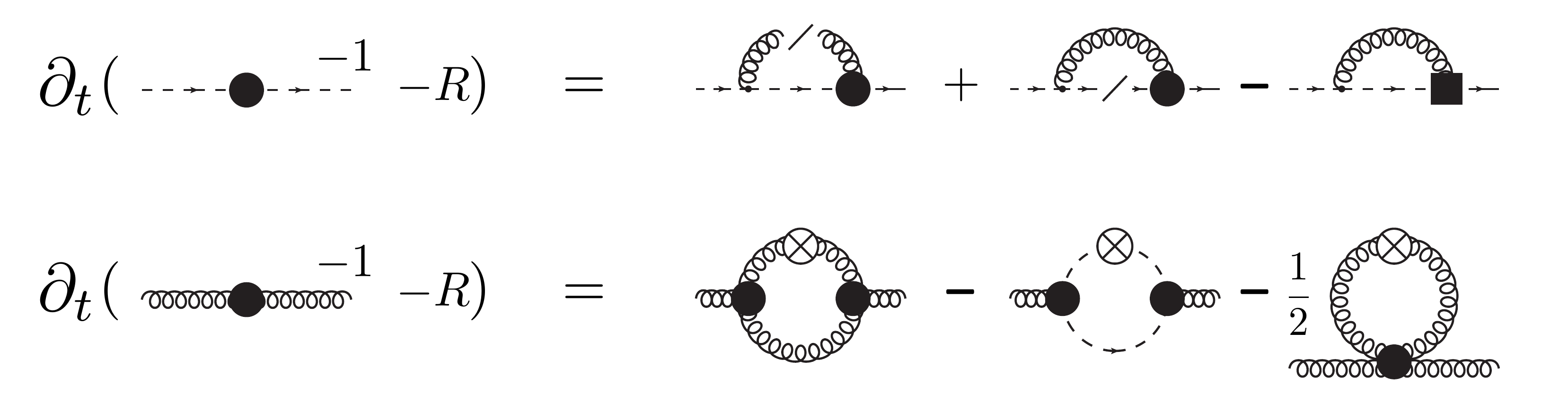}}\hfill\\
  \subfigure[Flow for the ghost-gluon vertex in the given 
truncation.]{\label{fig:Trunc_vertex}\includegraphics[width=0.65
\textwidth]{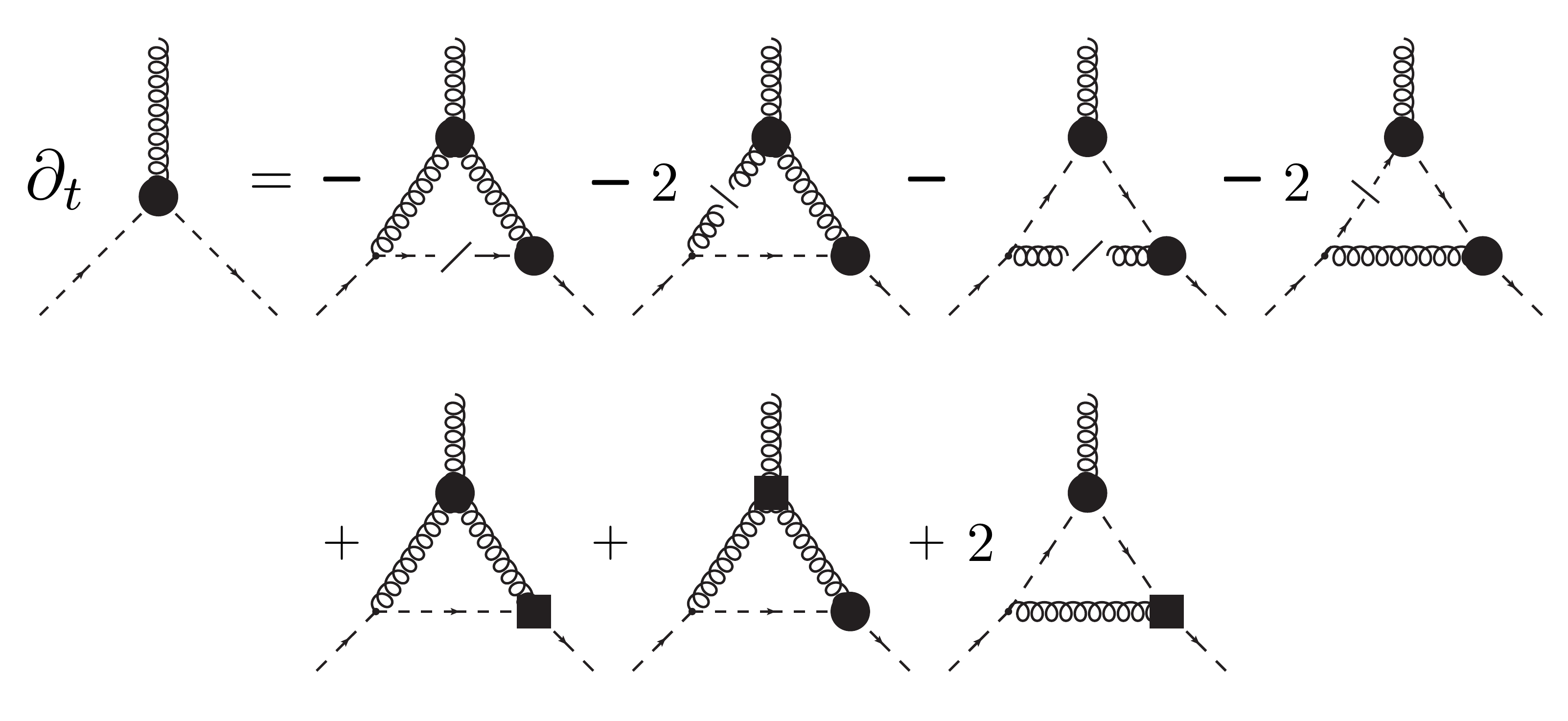}}\hfill\\
  \caption[]{Yang-Mills flows for propagators (a) and ghost-gluon
    vertex (b) the approximation discussed above. The flows for the
    ghost propagator and the ghost-gluon vertex are DSE-resummed and
    the ghost-tadpole in the gluon equation is neglected.}
\label{fig:truncation}
\end{center}
\end{figure}

The flow equations for the two-point functions depend on the two-,
three- and four-point functions. In particular we have tadpole
diagrams which depend on the ghost-ghost and ghost-gluon scattering
vertices $\Gamma^{(4)}_{\overline{c} c\overline{c} c}$ and
$\Gamma^{(4)}_{\overline{c} A^2 c}$. Albeit these vertices are absent
on the classical level they become relevant in the non-perturbative
regime.  They can be taken into account via a DSE resummation of the
vertices in the flow. This turns the flow into the total scale
derivative of the DSE, see Fig.~\ref{fig:Trunc_props}. This reflects
the fact that, in general, the flow equation is the differential form
of the Dyson-Schwinger equation. We remark that this usually would
introduce the necessity of renormalising the corresponding DSE. In the
present parameterisation this is avoided due to locality, for details
see \cite{finiteTYM}.  As a result the total scale derivative of the
DSE for the ghost two-point function in the presence of a regulator term
is equivalent to the flow equation, but is free of four-point
functions, for details see \cite{finiteTYM}. The scale derivative
acting on the dressed propagator gives
\begin{equation}
\label{eq:singlescaleG}
\partial_t G\left[ \phi\right] = - G\left[\phi \right]\cdot  
\partial_t \left( \Gamma^{(2)}\left[\phi \right] + R_{\phi}\right)\cdot G\left[\phi \right].
\end{equation}
For the vertices we introduce a parametrisation that naturally
captures the renormalisation group behaviour, for details see
\cite{Fischer:2009tn}. With the wave-function renormalisation of
the ghost and gluon we write
\begin{eqnarray} 
\Gamma^{(n)}(p_1,...,p_{n})= \prod_{i=1}^{n} 
Z^{1/2}_{\phi_i}(p_i)\, {\mathcal{T}}(p_1,...,p_{n})\,.
\label{eq:Gmnsol}\end{eqnarray} 
The $Z_{\phi}(p)$-factors carry the RG-scaling of the vertex as well
as the momentum dependence, thus they take into account potential kinematic
singularities. The ${\mathcal{T}}$ are renormalisation group invariant
tensors, which carry the canonical momentum dimension as well as the
tensor and colour structure. For the ghost propagator we have sketched
the idea how the ghost-ghost and ghost-gluon scattering vertices can
be absorbed via a DSE resummation. With a similar resummation in the flow
of the ghost-gluon vertex in Fig.~\ref{fig:cAc} we can fully resolve
its RG running at the symmetric point $p^2=k^2$. We use the
parametrisation
\begin{equation} 
 \mathcal{T}_{\overline{c}Ac,\mu}^{abc}(q,p) =z_{k,\overline{c} A c}\,\frac{1}{g} [S^{(3)}_{\overline{c}Ac}(q,p)
  ]^{abc}_\mu= z_{k,\overline{c} A c}
 \,i q_\mu f^{abc}\,, 
\end{equation}
where $g$ is the classical coupling. The RG invariant factor
$z_{k,\overline{c} A c}$ defines the running coupling
$\overline{\alpha}_s = z_{k,\overline{c} A c}^2/(4 \pi)$.  Also for
the three- and four-gluon vertices we restrict ourselves to the
classical tensor and colour structure. The parametrisation is done
according to eq. (\ref {eq:Gmnsol}), with
\begin{equation}
\mathcal{T}_{A^3} = z_{k,A^3}\,\frac{1}{g} S^{(3)}_{A^3} \,, 
\qquad \mathcal{T}_{A^4} = z_{k,A^4}\,\frac{1}{g^2} S^{(4)}_{A^4} \,.
\end{equation}
For large momenta the couplings $ z_{k,A^3},\ z_{k,A^4}$ relate to the
ghost-gluon dressing $z_{k,\overline{c} A c}$. In the infrared the
couplings are suppressed strongly. This is taken into account via
functions that approach $z_{3} = z_{4}=1$ for $k\gg \Lambda_{QCD}$,
but suppress the gluonic vertices in the infrared. The couplings are
given by $z_{k,A^3} = z_3\, z_{k,\overline{c} A c}, \ z_{k,A^4} =
z_4\, z_{k,\overline{c} A c}^2$. In addition to that we regularise the
vertex prefactors in such a way that we freeze the $Z_{A}$ for scales
smaller than the minimal turning point in the wave-function
renormalisation. The naive ansatz of the pure wave-function
renormalisation would yield unphysical IR divergent flows, which would
spoil the IR suppression of the purely gluonic vertices.  The above
approximation scheme is depicted in Fig.~\ref{fig:truncation},
including the DSE resummations.

\section{Results for Propagators and the Ghost-Gluon Vertex}
\begin{figure}[t!]
  \subfigure[Zero mode of the longitudinal
  propagator.]{\label{fig:GL}\includegraphics[width=0.49\textwidth]{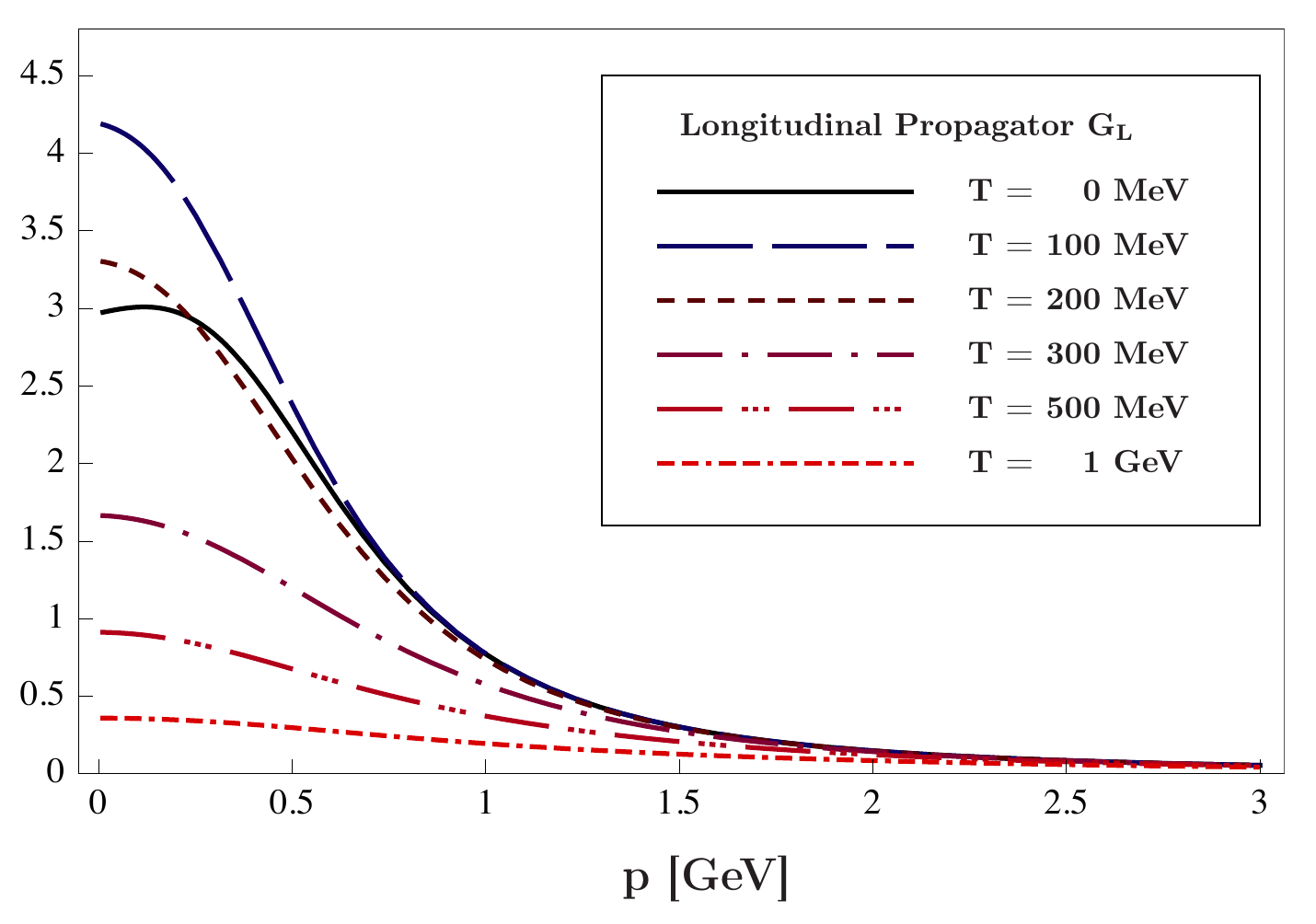}}\hfill
  \subfigure[Zero mode of the transverse propagator.]{\label{fig:GT}\includegraphics[width=0.49\textwidth]{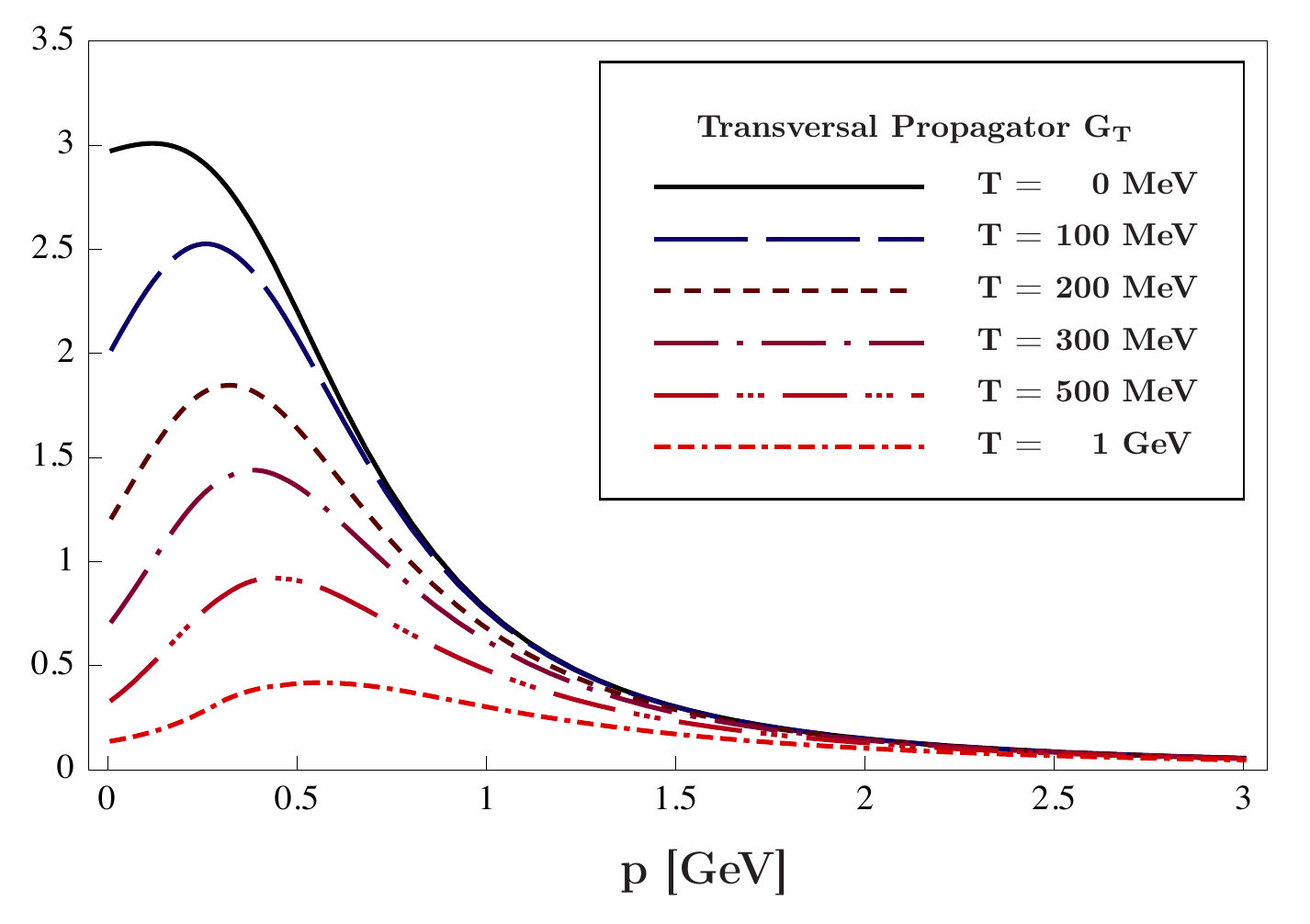}}\\
  \subfigure[Zero mode of the ghost wave function
  renormalisation.]{\label{fig:ZC}\includegraphics[width=0.49\textwidth]{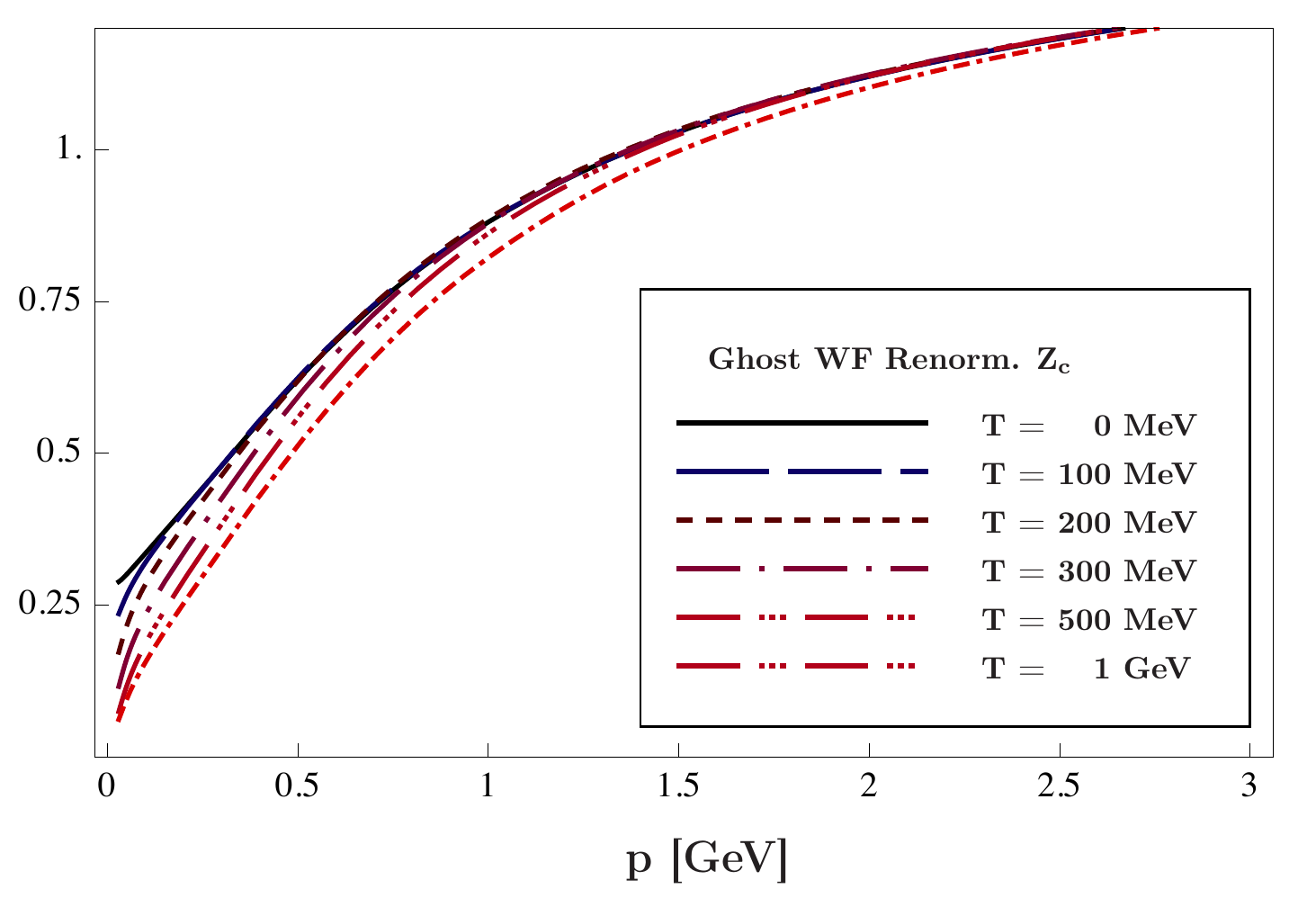}}\hfill
  \subfigure[Zero mode of the ghost propagator.]{\label{fig:GC}\includegraphics[width=0.49\textwidth]{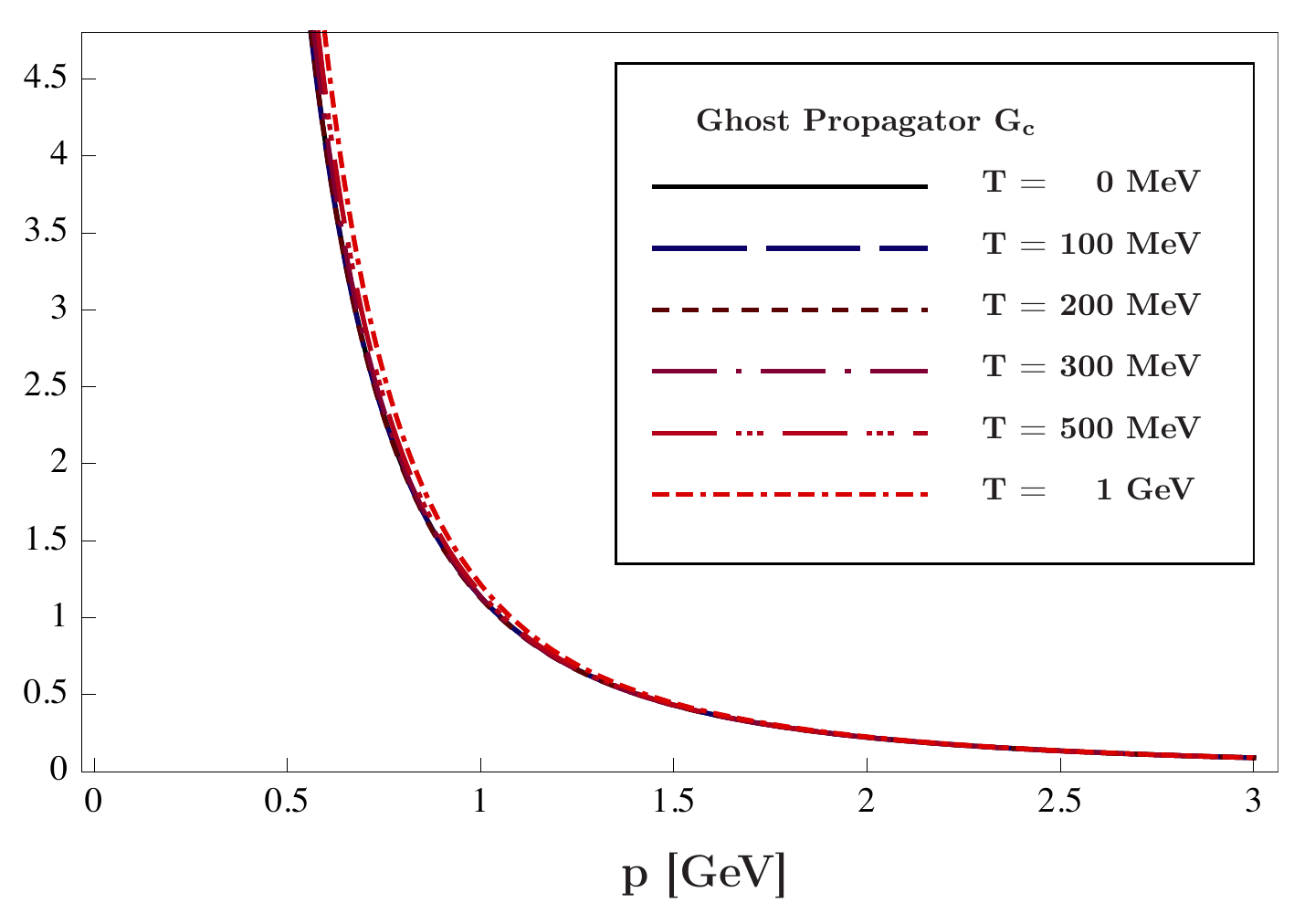}}\\
\begin{center}\subfigure[Ghost-gluon vertex coupling.]{\label{fig:ZcAc}\includegraphics[width=0.55\textwidth]{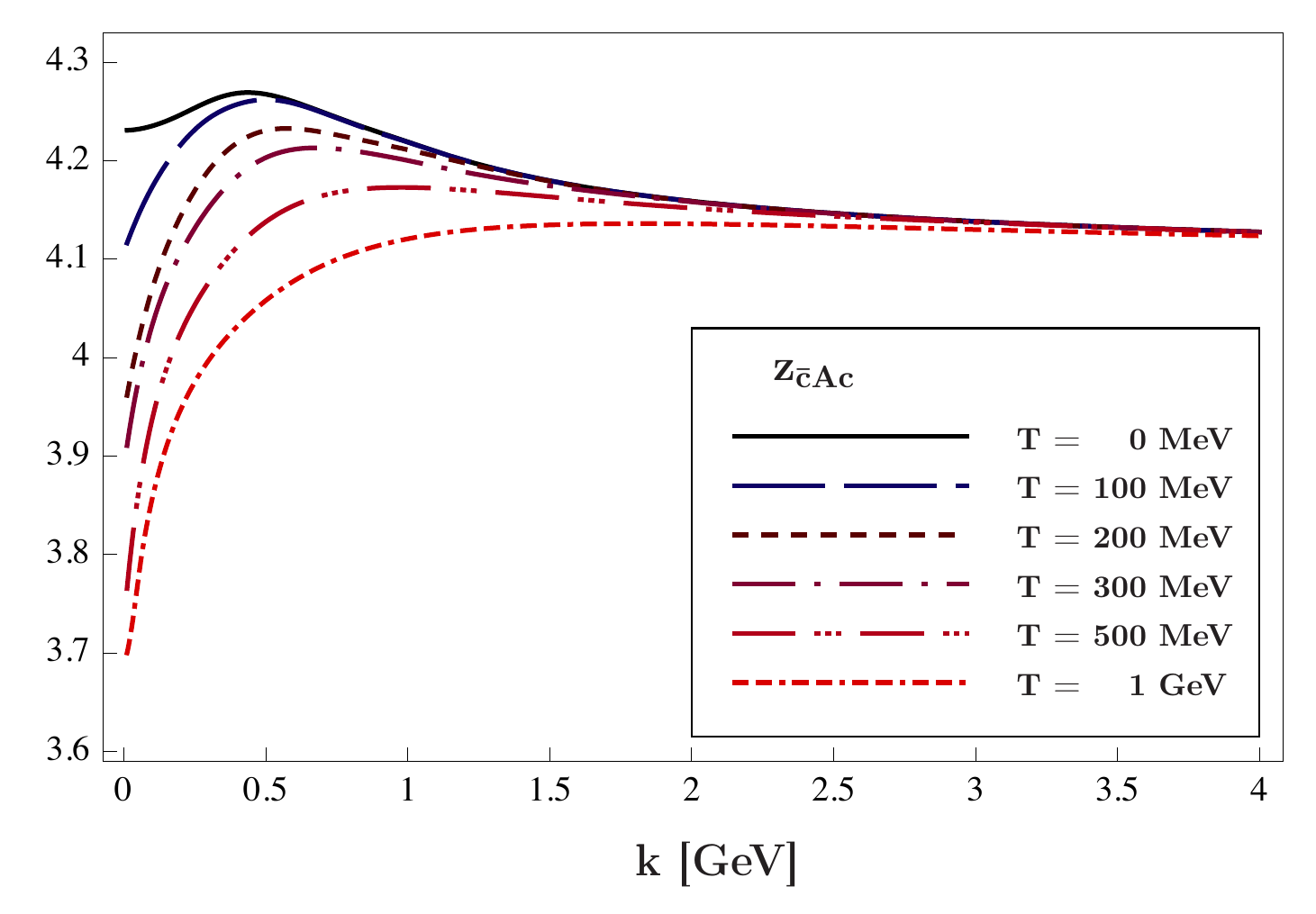}}
\end{center}
\caption[]{Yang-Mills propagators and the ghost-gluon vertex at
  various temperatures.}
\end{figure}
In this section we present the results for the ghost and gluon
propagators and the ghost-gluon vertex. The temperature is given in lattice units. 
Only below the typical temperature scale of $ 2 \pi T$
we have significant thermal effects on the momentum dependence. In
turn, above this scale the temperature fluctuations are suppressed and
all propagators tend towards their vacuum counterparts at vanishing
temperature. This also holds true for the ghost-gluon vertex. 

The most significant effect can be seen for the chromoelectric and
chromomagnetic gluon propagators, the components of the propagator
longitudinal and transversal to the heat bath respectively.  The zero
mode of the longitudinal gluon propagator at various temperatures is
given in Fig.~\ref{fig:GL} as a function of spatial momentum.

For low temperatures $T\lesssim 150\textnormal{ MeV} $ we see an
enhancement of the longitudinal propagator. Such an enhancement is
also seen on the lattice,
\cite{Maas:2011se,Cucchieri:2007ta,Fischer:2010fx,Bornyakov:2011jm,Aouane:2011fv,Maas:2011ez,Cucchieri:2011di}.  
For higher temperatures the longitudinal propagator is suppressed
relative to the gluon propagator at vanishing temperature.  This is
the expected behaviour caused by the Debye screening mass due to the
thermal screening of the chromoelectric gluon. For asymptotically high
temperatures $\gg T_c$ the chromoelectric gluon decouples. The onset
of this behaviour at about $T\approx 150\textnormal{ MeV}$ is earlier
as in the respective lattice computations
\cite{Maas:2011se,Cucchieri:2007ta,Fischer:2010fx,Bornyakov:2011jm,Aouane:2011fv,Maas:2011ez,Cucchieri:2011di}, 
where the thermal decoupling takes place for temperatures larger than
the critical temperature.

In order to quantitatively capture this behaviour we have to extend
our present truncation with a self-consistent inclusion of the
Polyakov loop background as well as a better resolution of the purely
gluonic vertices for momenta and frequencies below $\Lambda_{\rm
  QCD}\propto T_c$. Above the confinement-deconfinement scale we see
quantitative agreement with the lattice results.

The transversal mode is not enhanced for small temperatures in clear
distinction to the longitudinal mode. It is monotonously decreased
with temperature, see Fig.~\ref{fig:GL}.  Moreover, it develops a
clear peak at about 500 MeV. This can be linked to positivity
violation which has to be present for the transversal mode as in the
high temperature limit it describes the remaining dynamical gluons of
three-dimensional Yang-Mills theory in the Landau gauge. The infrared
bending is more pronounced as that of respective lattice results,
which already holds at vanishing temperature. It is another choice for
the decoupling solution as discussed in
\cite{Fischer:2008uz}. Moreover, its strength may also depend on the
quantitative precision achieved for the gluonic vertices at these
scales. In turn, for larger momenta the transversal propagator agrees
well with the respective lattice propagator.

The ghost shows only a small temperature dependence in
contradistinction to the gluonic propagators. The temperature
dependence is hardly evident on the level of the propagator
Fig.~\ref{fig:GC}, but can be resolved on the level of the
wave-function renormalisation Fig.~\ref{fig:ZC}. The wave-function
renormalisation is slightly suppressed, which corresponds to a
successive enhancement of the ghost propagator at finite temperature.
The enhancement of the ghost propagator is potentially self-amplifying
as it feeds back into the flow of the ghost two-point function, see
Fig.~\ref{fig:truncation}. This would
cause a pole in the ghost propagator at some temperature if the
ghost-gluon vertex would not change. However, in this case the flow of
the latter is dominated by the ghost diagram, see
Fig.~\ref{fig:cAc}. This non-trivial interplay of the flow of the ghost propagator with
the flow of the ghost-gluon vertex leads to a self-stabilising system
and prevents a further enhancement of the ghost.  This effect is
crucial for the stability of the solution of the Yang-Mills system at
finite temperature. Indeed, for a constant vertex no solution could be
obtained for intermediate temperatures $T\approx T_c$ and above. Thus
we conclude that any reliable truncation must comprise direct thermal
effects also in $n$-point functions with $n\leq3$.

The self-stabilising property of the Yang-Mills system explained above
is clearly seen in the temperature-dependence of the ghost-gluon
vertex.  The vertex is suppressed successively with the temperature,
see Fig.~\ref{fig:ZcAc}, which in turn ensures the relatively mild
change of the ghost propagator. Especially the sharp drop-off of the
vertex at small scales $k$ accounts for the smallness of the thermal
fluctuations to the ghost propagator.

\begin{figure}[t!]
  \subfigure[Comparison of the longitudinal propagator from above with
  the corresponding lattice
  result.]{\label{fig:FRG_latt_long}\includegraphics[width=0.49
    \textwidth]{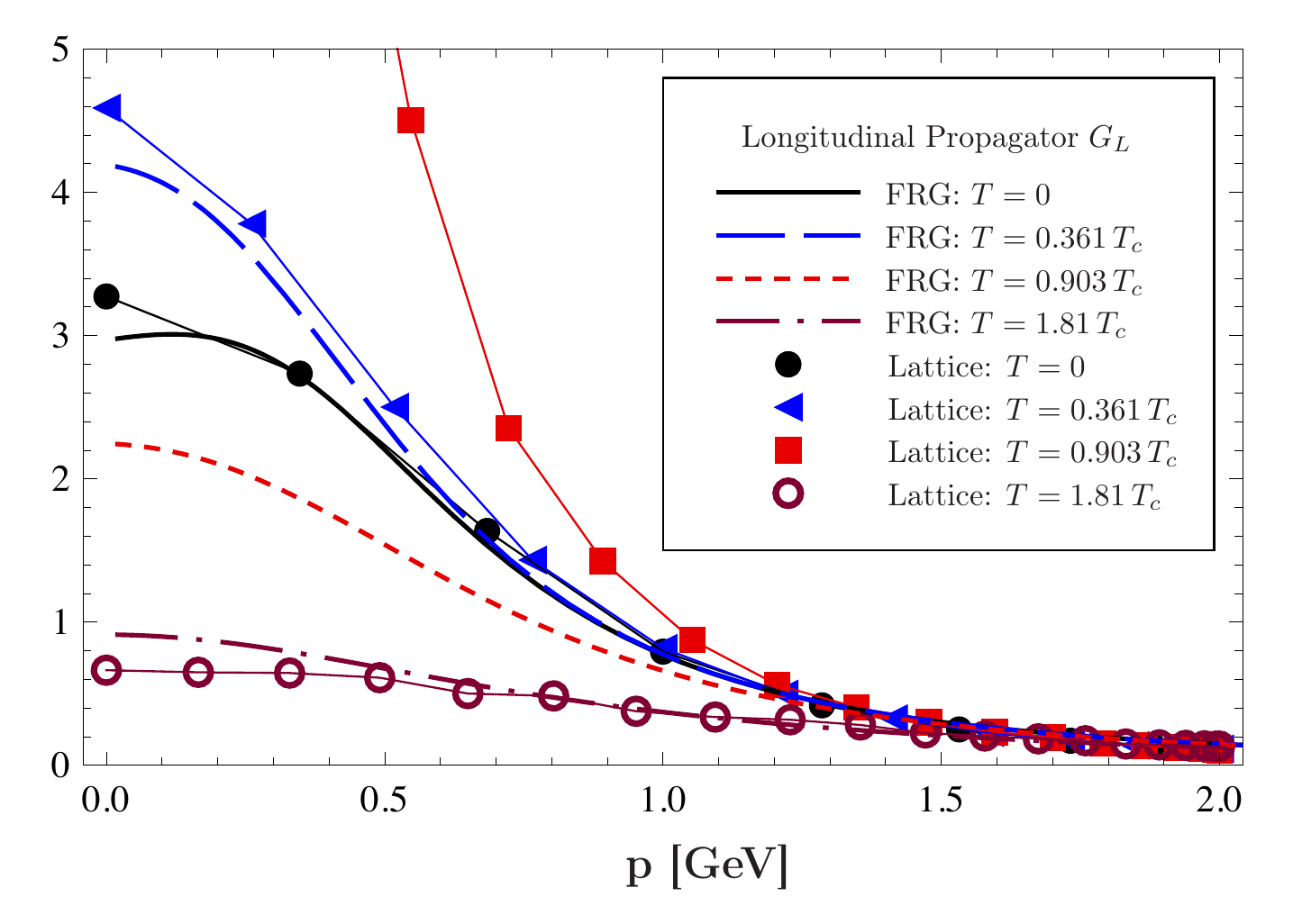}}\hfill \subfigure[Comparison
  of the transversal propagator from above with the corresponding
  lattice result.]{\label{fig:FRG_latt_trans}
\includegraphics[width=0.49\textwidth]{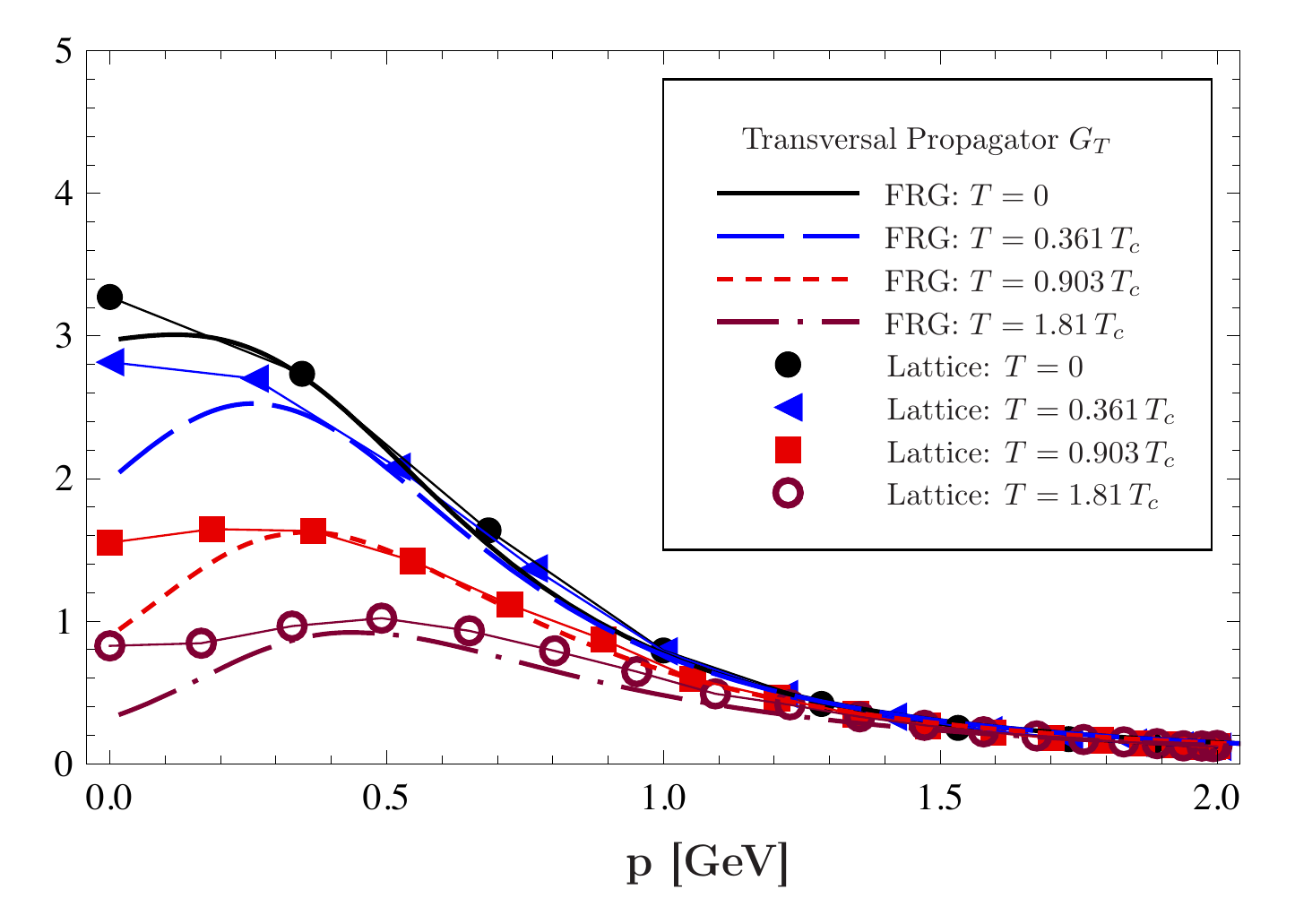}}\\
\caption{Comparison of the gluon propagator with the lattice results} 
\end{figure}

In the following, we compare the propagators above with lattice
results \cite{Fischer:2010fx}. For this purpose, we scale the lattice
data such that the lattice propagators at vanishing temperature match
our normalisation at momenta $p\gtrsim 1\textnormal{ GeV}$. Take
notice, that we did not use the lattice propagator as the initial
condition, thus the deep infrared of the data deviates from our
propagator already at zero temperature, which persists also in the
propagators at finite temperature. Apart from that, there is
quantitative agreement with the lattice data with respect to the
(temperature dependent) momentum region, where the thermal effects
appear. In Fig.~\ref{fig:FRG_latt_trans} the transversal propagators
are compared.

Clearly, we match the lattice propagator, except for the strong
bending of the FRG propagator in the infrared region. As already
discussed above, this difference is a direct consequence of the
deviation of ghost and gluon propagators at vanishing temperature in
the deep infrared, and the ansatz for the gluonic vertices. However,
the behaviour of the magnetic gluon agrees with the lattice
observations for momenta above 500 MeV and all temperatures. In
contrast to this, the electric gluon, see Fig.~\ref{fig:FRG_latt_long}
on the lattice shows a qualitatively different behaviour for
temperatures below and around the phase transition. Note also that
although the longitudinal propagators agree for $T=0.361T_c\approx
100\textnormal{MeV}$, it is exactly this region where the uncertainty
due to the truncation for the gluonic vertices is potentially large,
for a detailed discussion see \cite{finiteTYM}.
Being aware of a possible truncation dependence in the deep infrared
of the longitudinal propagator at low temperatures, we note that in
the present truncation the electric gluon qualitatively shows the
enhancement found on the lattice. Increasing the temperature this
feature disappears, and we see a qualitatively different effect for
temperatures below $T_c$. While the continuum result shows a strictly
monotonic decreasing propagator, the counterpart on the lattice is
enhanced in the confining regime, but reflects the phase transition in
form of a rapid decrease at $T_c$. Nevertheless, this deflection is
expected to be missed in the present truncation, as the full Polyakov
loop potential $V(A_0)$, see \cite{Braun:2007bx,Braun:2010cy}, is
pivotal for the critical behaviour around the phase transition, see
\cite{Maas:2011ez}. In a full calculation the inverse propagator is
proportional to the second derivative of the Polyakov loop
$\Gamma_{A,L}^{(2)}\sim V''(A_0)$, however in the work presented here
this term was dropped. In any case this term is absent in the magnetic
modes.

\section{Summary and Outlook}
We have computed temperature dependent Yang-Mills propagators in
Landau gauge in the framework of the functional renormalisation group
for $T\lesssim 3 T_c$. For this purpose we employ thermal flows, which
encode thermal fluctuations in the difference of the full
renormalisation group flow at non-vanishing and zero temperature.

The chromoelectric propagator shows the expected Debye-screening for
$T>T_c$ in quantitative agreement with the lattice results. For small
temperatures it shows qualitatively the enhancement also seen on the
lattice
\cite{Maas:2011se,Cucchieri:2007ta,Fischer:2010fx,Bornyakov:2011jm,Aouane:2011fv,Maas:2011ez,Cucchieri:2011di}.
However, the significance of the lattice results so far as well as
quantitative details are not settled yet. This concerns in particular
the behaviour of the chromoelectric propagator at criticality, see
\cite{Maas:2011ez}. We hope to add something to the clarification of
this issue within an extension of the present work.  There we resolve
the dependence on the Polyakov loop as well as have a better grip on
the gluonic vertices. The chromomagnetic propagator shows the expected
thermal scaling and tends towards the three-dimensional gluon
propagator in quantitative agreement with the lattice.

The ghost propagator only shows a mild enhancement with temperature in
agreement with the lattice. In contradistinction we see a strong
thermal infrared suppression of the ghost-gluon vertex which increases
with temperature.

At present we improve on the approximations and compute thermodynamic
observables, e.g. the pressure. Furthermore, we extend our work to
full QCD with 2 and 2+1 flavours in an extension of
\cite{Braun:2009gm} also at finite density.

\bibliographystyle{bibstyle}
\bibliography{bib}

\end{document}